\documentclass[twocolumn,amsmath,amssymb,aps]{revtex4}

\usepackage{graphicx}% Include figure files

\begin{document}

\title{A neural mechanism for binaural pitch perception via ghost stochastic resonance}

\author{Pablo Balenzuela}
\email{pablo.balenzuela@upc.es} \altaffiliation[also at ]{Fundaci\'on Antorchas, Chile 300 (1098), Buenos Aires,
Argentina.}
\author{Jordi Garc\'ia-Ojalvo}
\email{jordi.g.ojalvo@upc.es}
%\author{...}
\affiliation{ Departament de F\'isica e Enginyeria Nuclear, Universitat Polit\`ecnica de Catalunya, Colom 11,
E-08222 Terrassa, Spain }

%\author{Pablo Balenzuela}
 %\homepage{http://www.Second.institution.edu/~Charlie.Author}
%\affiliation{
%Fundacion Antorchas, Chile 300 (1098), Buenos Aires, Argentina.
%\textbackslash\textbackslash
%}

\date{\today}

\begin{abstract}
We present a physiologically plausible binaural mechanism for the perception of the pitch of complex sounds via
ghost stochastic resonance. In this scheme, two neurons are driven by noise and a different periodic signal each (with frequencies $f_1=kf_0$ and $f_2=(k+1)f_0$, where $k>1$), and their outputs (plus noise) are applied synaptically to a third neuron. Our numerical results, using the Morris-Lecar neuron model with chemical synapses
explicitly considered, show that intermediate noise levels enhance the response of the third neuron at
frequencies close to $f_0$, as in the cases previously described of ghost resonance. For the case of a inharmonic
combination of inputs ($f_1=kf_0+ \Delta f$ and $ f_2=(k+1)f_0 + \Delta f$) noise is also seen to enhance the rates of most probable spiking for the third neuron at a frequency $f_r = f_0 + \frac{\Delta f}{k+1/2}$.
In addition, we show that similar resonances can be observed as a function of the synaptic time constant. The
suggested ghost-resonance-based stochastic mechanism can thus arise either at the peripheral level or at a higher level of neural processing in the perception of pitch.
\end{abstract}

%\pacs{}

%\keywords{Suggested keywords}
%Use showkeys class option if keyword display desired

\maketitle

\noindent {\bf  The perception and processing of environmental complex signals resulting from the combination of
multiple inputs is a nontrivial task for the nervous system. In many species, solving efficiently this sensory
problem could have an evolutionary payoff. A classical example is the perception of the pitch of complex sounds
by the auditory system, the mechanism of which remains controversial.  Recently, a mechanism for the perception of
pitch has been proposed on the basis of the so-called ghost stochastic resonance. Under this paradigm, an
appropriate level of noise yields an optimal subharmonic neural response to a combination of two or more
harmonic signals that lack the fundamental frequency, which is nevertheless perceived by the system.
The original proposal concentrated in the peripheral level of the perception process, by considering the case of a simple monoaural presentation of the complex signal. On the other hand, it is known that complex sounds are also perceived when its two constituent tones are presented binaurally ({\em i.e.} one in each ear). Thus, the question that remains is whether ghost stochastic resonance can participate in detecting this ``virtual'' dichotic pitch at a higher level of processing. In this paper we present, on the basis of numerical simulations, a plausible mechanism for the binaural perception of the pitch
of complex signals via ghost stochastic resonance. In this scenario, each of the two input tones drives a
different noisy neuron (corresponding to detection in the left/right auditory pathways), and together they drive a
third noisy neuron that perceives the missing fundamental. In this way, the same basic mechanism of ghost
resonance can explain pitch perception occurring at both the peripheral and a higher processing level.}

\section{\label{sec:intro}Introduction}

\subsection{\label{sec:single}Pitch perception by single neurons}

Under many conditions sensory neurons can be considered as noisy threshold detectors, responding to external
signals (either from the environment or from other neurons) in an all-or-none manner. Substantial effort has
been dedicated to examine theoretically and numerically the response of neurons to simple input signals, usually
harmonic, both under deterministic \cite{punit,andre1} and stochastic \cite{dante0,physrep} conditions.

Much less studied is the case of multiple input signals. It is known, for instance, that a high frequency signal
enhances the response of a neuron to a lower frequency driving via vibrational resonance \cite{vibr}. On the
other hand, two-frequency signals are commonly used for diagnostic purposes, such as in the analysis of evoked
potentials in the human visual cortex \cite{cortex}, but the detection and processing of this type of combined
signals is poorly understood. Recently, a study of the response of a neuron to a combination of harmonics in
which the fundamental is missing \cite{dante1} has shed new light upon the problem of the perception of the
pitch of complex sounds \cite{dante2}.

The perceived pitch of a pure tone is simply its frequency. In contrast, the perceived pitch of a complex sound
(formed by a combination of pure tones) is a subjective attribute, which can nevertheless be quantified accurately
by comparing it with a pure tone. In the particular case of harmonic complex sounds (signals whose constituent
frequencies are multiple integers of a fundamental frequency), the perceived pitch is the fundamental, even
if that frequency is not spectrally present in the signal. For that reason, the pitch is usually referred to in
this case as a ``virtual pitch'', and its perception is sometimes called the ``missing fundamental illusion''.

The neural mechanism underlying pitch perception remains controversial. From a neurophysiological perspective,
the perceived pitch is associated with the inter-spike interval statistics of the neuronal firings
\cite{cariani1,cariani2}. The analysis presented in Refs. \cite{dante1,dante2} shows that a neuron responds
optimally to the missing fundamental of a harmonic complex signal for an intermediate level of noise, making use
of two ingredients: (i) a linear interference of the individual tones, which naturally leads to signal peaks at
the fundamental frequency, and (ii) a nonlinear threshold that detects those peaks (with the help of a suitable
amount of noise, provided the signal is deterministically subthreshold). The behavior of this relatively simple
model yields remarkably good agreement with previous psychophysical experiments \cite{schouten}. The phenomenon
has been termed {\em ghost stochastic resonance} (GSR), and has been replicated experimentally in excitable
electronic circuits \cite{oscar} and lasers \cite{ghost}.

\subsection{\label{sec:binaural}Signal integration and processing of distributed inputs}

Besides the question of {\em how} pitch is perceived, another contested debate relates to {\em where} perception
takes place. Although interval statistics of the neuronal firings \cite{cariani1,cariani2} show that pitch
information exists in peripheral neurons, other results seem to indicate that, at least to some extent, pitch perception takes place at a
higher level of neuronal processing \cite{pantev}. A typical example is found in binaural experiments, in which two
components of a harmonic complex signal enter through different ears. It is known that in that case a (rather
weak) low-frequency pitch is perceived. This is called ``dichotic pitch'', and can also arise from the binaural
interaction between broad-band noises. For example, Cramer and Huggins \cite{cramer} studied the effect of a
dichotic white noise when applying a progressive phase shift across a narrowband of frequencies, centered on 600
Hz, to only one of the channels. With monaural presentation listeners only perceived noise, whereas when using
binaural presentation over headphones, listeners perceived a 600-Hz tone against a background noise.

It is worth examining whether the ghost resonance mechanism introduced by Chialvo {\em et al.}
\cite{dante1,dante2} can also account for the binaural effects described above. Ghost resonance has already been
seen to be enhanced by coupling in experiments with diffusively coupled excitable lasers \cite{bghost}, but no
studies in synaptically coupled neurons have been made so far. Given that chemical synapses lead to pulse
coupling, a reliable coincidence detection is required in order for ghost resonance to arise in this case. In
what follows we examine the situation in which two different neurons receive one single component of the
complex signal each (so that each neuron represents detection at a different auditory channel in a binaural
presentation), and act upon a third neuron which is expected to perceive the pitch of the combined signal. Our
results show that this higher-level neuron is indeed able to perceive the pitch, hence providing a neural
mechanism for the binaural experiments.

\section{Model description}

\subsection{Neuron Model}

We describe the dynamical behavior of the neurons with the Morris-Lecar model \cite{morlec},
\begin{eqnarray}
\frac{dV_i}{dt} & = & \frac{1}{C_m}(I^{\rm app}_i - I^{\rm ion}_i-I^{\rm syn}_i)+D_i\xi(t) \label{eqV} \\
\frac{dW_i}{dt} & = & \phi \Lambda(V_i)[W_{\infty}(V_i) - W_i]   \label{eqW}
\end{eqnarray}
where $V_i$ and $W_i$ stand for the membrane potential and the fraction of open potasium channels, respectively,
and the subindex $i$ labels the different neurons, with $i=1,2$ representing the two input neurons and $i=3$
denoting the processing neuron. $C_m$ is the membrane capacitance per unit area, $I^{\rm app}_i$ is the external
applied current, $I^{\rm syn}_i$ is the synaptic current, and the ionic current is given by
\begin{eqnarray}
&&I^{\rm ion}_i  =  g_{Ca}M_{\infty}(V_i)(V_i-V^0_{Ca})+ \nonumber \\
 & & \qquad\qquad\qquad g_KW_i(V_i-V^0_K)+g_L(V_i-V^0_L) \label{eqIion}
\end{eqnarray}
where $g_a$ ($a=Ca,K,L$) are the conductances and $V^0_a$ the resting potentials of the calcium, potassium and
leaking channels, respectively. The following functions of the membrane potential are also defined:
\begin{eqnarray}
&& M_{\infty}(V) =  \frac{1}{2}\left[1+\tanh\left(\frac{V-V_{M1}}{V_{M2}}\right)\right]      \\
&&W_{\infty}(V) = \frac{1}{2}\left[1+\tanh\left(\frac{V-V_{W1}}{V_{W2}}\right)\right]      \\
&&\Lambda(V) = \cosh\left(\frac{V-V_{W1}}{2V_{W2}}\right)\,,
\end{eqnarray}
where $V_{M1}$, $V_{M2}$, $V_{W1}$ and $V_{W2}$ are constants to be specified later. The last term in Eq.
(\ref{eqV}) is a white noise term of zero mean and amplitude $D_i$, uncorrelated between different neurons.

In the deterministic and single-neuron case, this system shows a bifurcation to a limit cycle for increasing
applied current $I^{\rm app}$ \cite{bifutrue}. This bifurcation can be a saddle-node (type I) or a subcritical Hopf
bifurcation (type II) depending on the parameters. We chose this last option for the numerical calculations
presented in this paper. The specific values of the parameters used in what follows are shown
in table (\ref{tab:ML}) \cite{bifu}.
The equations were integrated using the Heun method \cite{nises}, which is equivalent to
a second order Runge-Kutta algorithm for stochastic equations.

\subsection{Synapsis model}

In this work we couple the neurons using a simple model of chemical synapsis \cite{destexhe}. In this model, the
synaptic current through neuron $i$ is given by
\begin{equation}
I^{\rm syn}_i = \sum_{j\in {\rm neigh}(i)} g_i^{\rm syn}r_j(V_i-E_s), \label{syn}
\end{equation}
where the sum runs over the neighbors that feed neuron $i$, $g^{\rm syn}_i$ is the conductance of the
synaptic channel, $r_j$ stands for the fraction of bound receptors of the postsynaptic channel, $V_i$ is the
postsynaptic membrane potential, and $E_s$ is a parameter whose value determines the type of synapsis (if larger
than the rest potential, {\em e.g.} $E_s=0$~mV, the synapsis is excitatory; if smaller, {\em e.g.} $E_s=-80$~mV,
it is inhibitory).

The fraction of bound receptors, $r_i$, follows the equation
\begin{equation}
\frac{dr_i}{dt}= \alpha [T]_i(1-r_i) - \beta r_i\,,
\end{equation}
where $[T]_i=\theta(T_0^i +\tau_{\rm syn}-t)\theta(t-T_0^i)$ is the concentration of neurotransmitter released
in the synaptic cleft, $\alpha$ and $\beta$ are rise and decay time constants, respectively, and $T_0^i$ is the
time at which the presynaptic neuron (labeled now $i$) fires, what happens whenever the presynaptic membrane
potential exceeds a predetermined value, in our case chosen to be $10$~mV. The time during which the synaptic connection is active is given by $\tau_{\rm
syn}$. The values of the parameters that we use, specified below, were taken from \cite{destexhe} and could be read from table (\ref{tab:ML}). 

\begin{table}[htbp]
\begin{center}
\begin{minipage}[t]{2.5in}
\centering
\begin{tabular}{|c|c|}
 \hline
  \textbf{Parameters}  & \textbf{Morris-Lecar: TII} \\ \hline
 $C_m$  &  $5\,\mu \mathrm{F/cm}^2$    \\ \hline
 $g_K$  &  $8\,\mu \mathrm{S/cm}^2$   \\  \hline
 $g_L$  &  $2\,\mu \mathrm{S/cm}^2$     \\ \hline
 $g_{Ca}$  & $4.4\,\mu \mathrm{S/cm}^2$    \\   \hline
 $V_K$  &  $-80\,\mathrm{mV}$     \\ \hline
 $V_L$  &  $-60\,\mathrm{mV}$     \\ \hline
 $V_{Ca}$ & $120\,\mathrm{mV}$     \\ \hline
 $V_{M1}$  &  $-1.2\,\mathrm{mV}$     \\ \hline
 $V_{M2}$  &  $18\,\mathrm{mV}$     \\ \hline
 $V_{W1}$  &  $2\,\mathrm{mV}$     \\ \hline
 $V_{W2}$  &  $30\,\mathrm{mV}$     \\ \hline
 $\phi$  & $1/25\, \mathrm{s}^{-1}$     \\ \hline \hline
  \textbf{Parameters}  & \textbf{Synapsis} \\ \hline
 $\alpha$  & $0.5\, \mathrm{ms}^{-1} \mathrm{mM}^{-1}$ 	\\ \hline
 $\beta$   & $0.1\, \mathrm{ms}^{-1}$ 	\\ \hline
 $g_{syn}$ & (specified in each case)	\\ \hline
 $\tau_{syn}$ & (specified in each case)  \\ \hline
 $E_s$ & $0\,\mathrm{mV}$	\\ \hline
 
\end{tabular}
\caption{Parameters values of the Morris-Lecar and synapse models used in this work.
\label{tab:ML}}
\end{minipage}
\end{center}
\end{table}

\section{The case of distributed harmonic complex signals}

As mentioned above, we consider a network of three neurons organized in two layers. The first layer is composed
of two units (called ``input neurons'') that receive the external inputs, and whose responses act upon the
processing layer, composed in this case of only one unit (called ``processing neuron'') . The coupling
is unidirectional from each of the input neurons to the processing neuron. Of course, physiological realism
dictates that more than three neurons will be present. However, we model here for simplicity the simplest possible case; one can expect that adding more neurons will only improve the results.

In order to analyze the global response of this network to a distributed complex signal, we apply to each one of
the input neurons a periodic external current with frequencies $f_1$ and $f_2$. In response these neurons emit a
sequence of spikes with inter-spike interval distributions centered at $f_1$ and $f_2$ and with variances
directly related to the noise intensities $D_1$ and $D_2$.

\begin{figure}
\includegraphics[height=5.8cm,keepaspectratio]{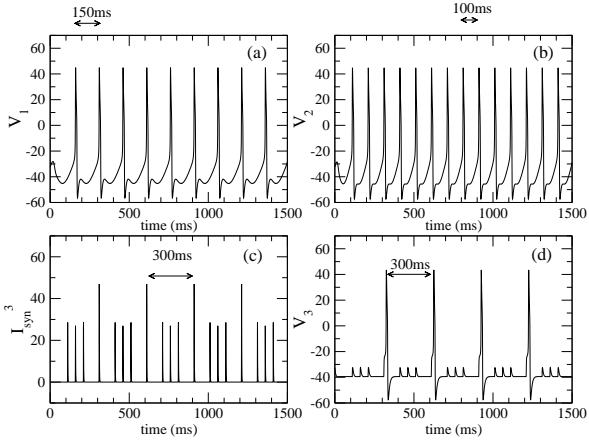}
\caption{\label{fig:noise0} Deterministic response to a distributed harmonic complex signal. The membrane
potential for the three neurons is shown: (a,b) input neurons, (d) processing neuron. The synaptic current
acting on neuron 3, $I_3^{\rm syn}$, is shown in plot (c). The two inputs neurons are fed with two sinusoidal
signals of periods $T_1=150$~ms and $T_2=100$~ms respectively (which gives a ghost resonance of $T_0=300$~ms).
The synaptic coupling between input and processing neurons is $g_{syn}=2$~nS and $\tau_{syn}=1$~ms. All noise intensities are zero $D_1=D_2=D_3=0$.}
\end{figure}

Figure \ref{fig:noise0} shows the behavior of the system in the absence of noise ($D_1=D_2=D_3=0$). In this
deterministic situation, the input neurons fire exactly with the frequencies at which they are modulated. If
$f_1=kf_0$ (or, equivalently, $T_1=T_0/k$, where $T_i$ is the period corresponding to the frequency $f_i$) and
$f_2=(k+1)f_0$ ($T_2=T_0/(k+1)$), the two input neurons exhibit simultaneous spikes at intervals $T_0=1/f_0$
(provided the two harmonic signals are in phase), so that the synaptic current acting on the third neuron has
maxima with the same frequency, as can be observed in Fig. \ref{fig:noise0}(c). In this example $T_1=100$~ms and
$T_2=150$~ms, so that $k=2$ and $T_0=300$~ms. Under these conditions, and for an adequate value of the coupling
strength $g_{\rm syn}$, the processing neuron fires with frequency $f_0=1/300$~kHz, as shown in Fig.
\ref{fig:noise0}(d). Hence, in this case a deterministic ghost resonance is observed.

The previous example, however, is not realistic, since in normal conditions a neuron is affected by a substantial
level of noise coming from, among other sources, the background activity of other neurons acting upon it. This
causes a drift in the spike times and a broadening in the distribution of inter-spike intervals. We will now
show that even in this case the missing fundamental frequency can be successfully detected, as was suggested in
\cite{dante1,dante2} for a single neuron, even if the synaptic coupling $g_{\rm syn}$ and the applied current in
the output neuron, $I_3^{\rm app}$, are slightly below the bifurcation threshold, so that the neuron does not
fire in absence of noise ($D_3=0$).

\begin{figure}
\includegraphics[height=5.8cm,keepaspectratio]{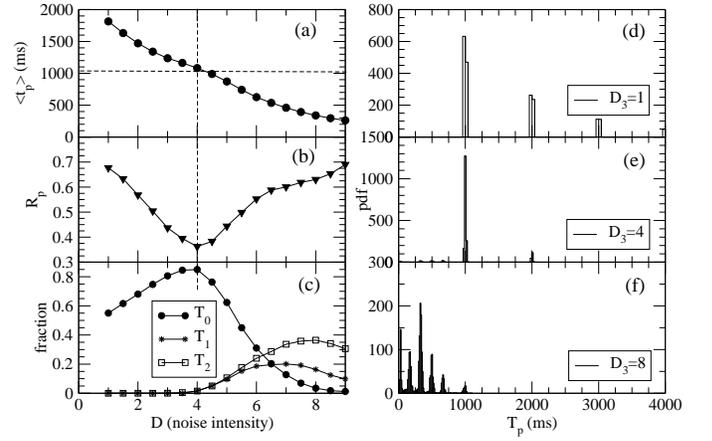}
\caption{\label{fig:tau35} Left panels: response of the processing neuron for increasing noise intensity: (a)
mean time between spikes $\langle T_p\rangle $, (b) normalized variance of the distribution $R_p=\sigma_p/\langle
T_p\rangle $, and (c) fraction of pulses spaced around $T_0=1/f_0$, $T_1=1/f_1$ and $T_2=1/f_2$ as a function of
the noise amplitude in the processing neuron, $D_3$. Right panels: probability distribution functions of the
time between spikes $T_p$ for three values of the noise intensity $D_3$: (d) $D_3=1$, (e) $D_3=4$, and (f)
$D_3=8$. Parameters are $\tau_{\rm syn}=35$~ms and  $g_{syn}=1$~nS for the synapsis and we used $f_1=2$~Hz and
$f_2=3$~Hz (which gives $f_0=1$~Hz) for the sinusoidal signals in the input neurons.}
\end{figure}

With this in mind, we conduct a series of numerical experiments looking for the occurrence of ghost stochastic
resonance. We choose $f_1=2$~Hz and $f_2=3$~Hz, so the ghost resonance should be located at $f_0=1$~Hz. As is
usual in neurophysiology, in order to quantify the behavior of the system, we follow the time between
consecutive spikes, $T_p$. In what follows, we analyze the first two moments of the distribution of $T_p$, its
mean value $\langle T_p\rangle $ and its normalized variance $R_p = \sigma_p/\langle T_p\rangle $. To estimate
the coherence of the output with the frequencies of interest, we also compute the fraction $f_{t0}$ of
inter-spike intervals in the neighborhood of $T_0=1/f_0$. The dependence of these variables (corresponding to
the processing neuron) on the noise intensity $D_3$ is shown in Fig. \ref{fig:tau35}. These results display a clear
resonance at $D_3\sim 4$. The normalized variance of the inter-spike interval distribution [Fig.
\ref{fig:tau35}(b)] exhibits a minimum when the spikes of the third neuron are spaced, on average, $\langle
T_p\rangle =1000$~ms [Fig. \ref{fig:tau35}(a)]. Additionally, around $80\%$ of the spikes are spaced $\pm 5\%$
around $T_0=1000$~ms for $D_3\sim 4$ [Fig. \ref{fig:tau35}(c)]. These results clearly indicate that noise enhances the response of the processing neuron at the frequency $f_0$, which is not present in the input neurons.

The right panels of Fig. \ref{fig:tau35} show the probability distribution functions of the inter-spike
intervals $T_p$ for three values of the noise in the processing neuron. For low noise intensities, the neuron
spikes most likely when two input spikes arrive together, but with randomly one or more of these coincidence
events is skipped. For this reason, the probability distribution function shows peaks centered at multiples of
$T_0$, as it usually happens in conventional stochastic resonance \cite{SR}. As the noise level increases skips
occurs less frequently, until an optimal noise for which almost all spikes occur every $T_0$, {\em i.e.} at the
missing fundamental frequency. For even larger noise intensities, spikes appear at the original input frequencies $f_1=1/T_1$ and $f_2=1/T_2$, as can be observed in Fig. \ref{fig:tau35}(f).

\begin{figure}
\includegraphics[height=5.8cm,keepaspectratio]{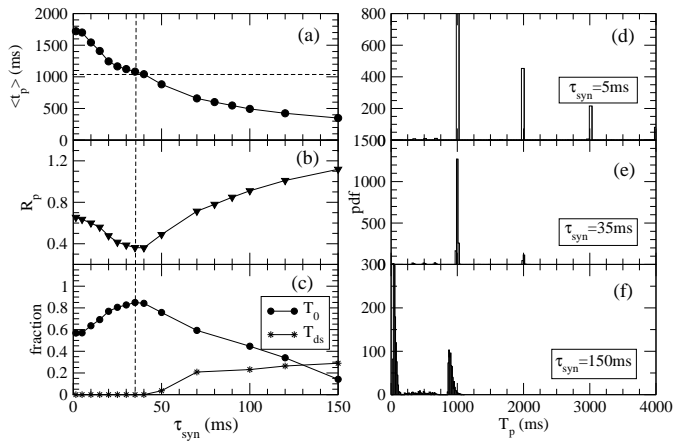}
\caption{\label{fig:tauvar} Left panels: (a) mean time between spikes, (b) normalized variance of the inter-spike
interval distribution, and (c) fraction of pulses around $T_0=1/f_0$ and $T_{\rm ds}=1/f_{\rm ds}$ as a function
of $\tau_{\rm syn}$. Right panels: Probability distribution functions of the inter-spike intervals $T_p$ for
three values of $\tau_{\rm syn}$: (d) $\tau_{\rm syn}=5$~ms, (e) $\tau_{\rm syn}=35$~ms, and (f) $\tau_{\rm
syn}=150$~ms. The value of $g_{syn}$ is different for each value of $\tau_{syn}$ ranging from $g_{syn}=2.50$~nS
for $\tau_{syn}=1.5$~ms up to $g_{syn}=1.00$~nS for $\tau_{syn}=150$~ms. $f_1=2$~Hz and $f_2=3$~Hz (which gives
$f_0=1$~Hz) for the sinusoidal signals in the input neurons was used.}
\end{figure}

In the binaural mechanism of ghost stochastic resonance described above, synaptic coupling obviously plays an
important role, since the transfer of the input modulation from the sensory neurons to the processing neuron
occurs synaptically. Taking into account that synaptic transmission is an intrinsically dynamical phenomenon
(whose temporal behavior we are modelling explicitly), it is natural to expect that the characteristic time
scale of this process will influence the occurrence of the resonance. Indeed, the results shown above correspond
to an optimal value of the synaptic time $\tau_{\rm syn}$.  As shown in Fig. \ref{fig:tauvar} for fixed noise
level $D_3$, a resonance in the response of the system to the missing fundamental is also observed with respect
to $\tau_{\rm syn}$.

We recall that $\tau_{\rm syn}$ represents the time during which the neurotransmitters remain in the synaptic
cleft before they start to disappear with rate $\beta$, and it is a measure of the width of the pulses of the
synaptic current received by the processing neuron. Therefore, for low $\tau_{\rm syn}$
[Fig.~\ref{fig:tauvar}(d)] the synaptic pulses are very narrow, and hence coincidence detection is compromised.
The characteristic probability distribution function in this case presents peaks at multiples of $T_0$,
indicating that even if the noise level is optimized, coincident spikes from input neurons are skipped.

\begin{figure*}[htb]
\includegraphics[height=5cm,keepaspectratio]{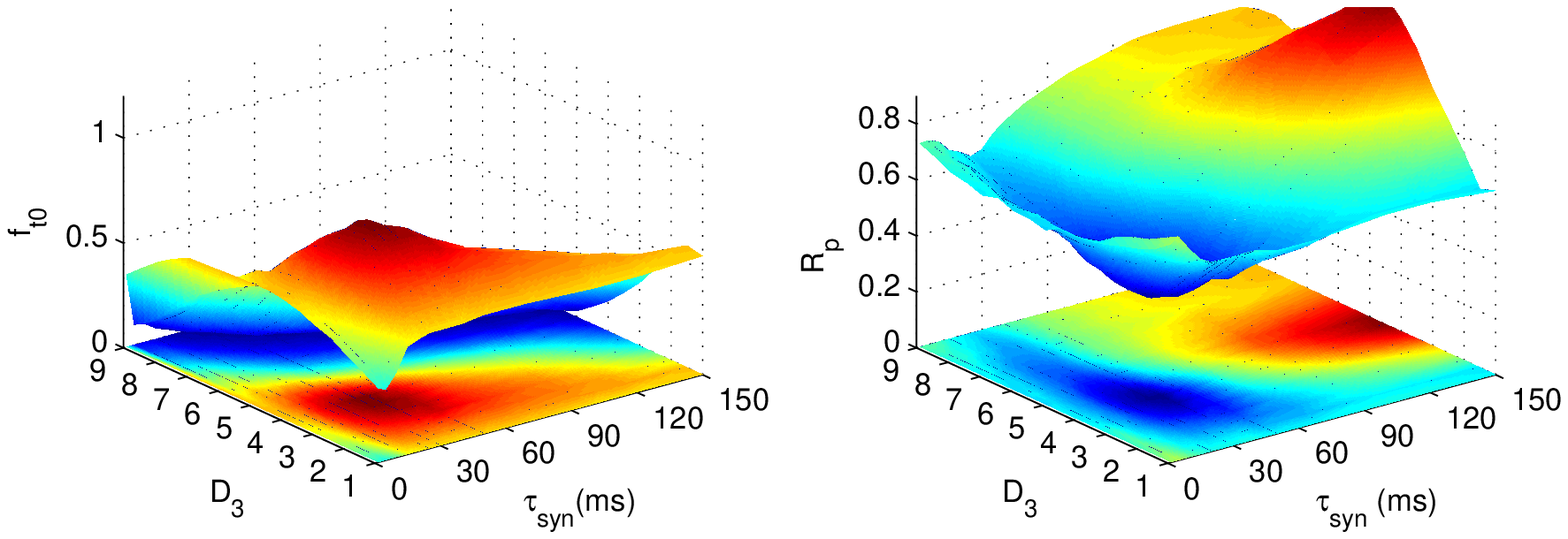}\caption{\label{fig:3D} Left: fraction of pulses $f_{t0}$ occurring at intervals $T_p$ equal ($\pm 5\%$) to the
period of the ghost resonance ($T_0=1/f_0$). Right: normalized variance of the inter-spike interval
distribution, $R_p$. Both quantities plotted as function of noise intensity ($D_3$) and $\tau_{\rm syn}$. }
\end{figure*}

As $\tau_{\rm syn}$ increases the current pulses widen and coincidence detection improves, so that an optimal
situation is reached for which the ghost resonance is very clear. But if we continue increasing the value of
$\tau_{\rm syn}$ the synaptic pulses become exceedingly wide and sequences of double spikes appear (spaced by
$T_{\rm ds}=1/f_{\rm ds}$). This happens because noise can excite two spikes while the synaptic current remains
high. Indeed, Fig.~\ref{fig:tauvar}(c)
shows that the fraction of spikes occurring at intervals around $T_{\rm ds}$ ($\pm5\%$) begins to be important
for $\tau_{\rm syn} > 50$~ms. The corresponding distribution function in Fig.~\ref{fig:tauvar}(f), shown here
for $\tau_{\rm syn}=150$~ms, corroborates this fact.

The joint effect of the synaptic time $\tau_{\rm syn}$ and the noise level $D_3$ can be observed in the
three-dimensional plots shown in Fig.~\ref{fig:3D}. This figure shows $R_p$ and $f_{t0}$ as a function of both
$D_3$ and $\tau_{\rm syn}$. We can see the response of the processing neuron at the missing fundamental is most
favorable when both parameters are simultaneously optimized. The normalized variance of the inter-spike interval
distribution, $R_p$, shows a clear minimum for $\tau_{\rm syn}\sim35$~ms and $D\sim4$. For these same parameter
values, the fraction of spikes $f_{t0}$ occurring at intervals around $T_0$ exhibits a maximum at almost $80\%$.

\section{The inharmonic case}

A paradigmatic experimental result in pitch perception refers to the pitch reported by human subjects to the
presentation of a inharmonic complex sound, in which the originally harmonic components of the
input are all shifted in frequency by a constant $\Delta f$. In such a way the individual component are still
separated in frequency by a constant missing ``fundamental'' $f_0$, but are no longer multiples of it. The
frequencies $f_1$ and $f_2$ are chosen to be
\begin{equation}
f_1=kf_0+\Delta f,  \qquad f_2=(k+1)f_0+\Delta f\,,
\end{equation}
with $k$ integer. In other words, $f_0$ is no longer the greatest common divider of $f_1$ and $f_2$, even though
it's still their difference. If the system is simply detecting the difference $f_2-f_1$, it should always
display a fixed resonance at $f_0$, independently of the frequency shift $\Delta f$. But if the pitch detection
does depends on $\Delta f$, it will no longer be perceived as the difference between the input frequencies. This
last situation is in fact what was found in human experiments \cite{schouten}. The neural mechanism proposed in
\cite{dante1,dante2} shows that the frequency of the ghost resonance shifts linearly with $\Delta f$ following
the relation,
\begin{equation}
f_r = f_0 + \frac{\Delta f}{k+1/2}\,. \label{eq:frf1}
\end{equation}
in agreement with the auditory experimental results of Refs. \cite{cariani1,cariani2,schouten}.

\begin{figure}
\includegraphics[height=5.8cm,keepaspectratio]{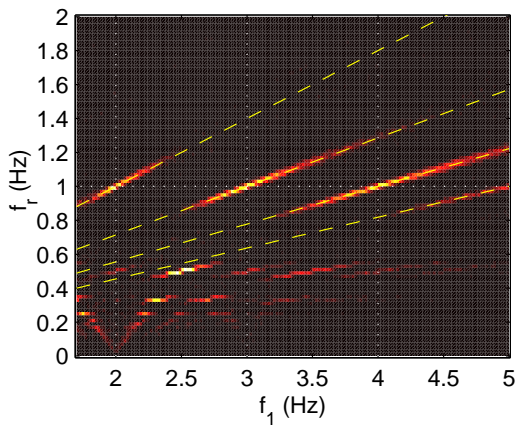}
\caption{\label{fig:frf1} Probability of observing a spike in the processing neuron with instantaneous rate $f_r$ (in gray scale) as a function of the frequency $f_1$ of one of the input neurons. We can observe a
remarkable agreement of the responses following the lines predicted by Eq.~(\ref{eq:frf1}) for $k=2,3,4,5$ (dashed lines from top to bottom).
Parameters: $\tau_{syn}=35$~ms, $g_{syn}=1.0$~nS, $D_3=4.0$, $f_1=2Hz+\Delta f$, $f_2=3Hz+\Delta f$.}
\end{figure}

We now examine whether a scaling similar to that of Eq.~(\ref{eq:frf1}) is observed in the response of the processing
neuron. We fix the noise intensity $D_3$ and synaptic time $\tau_{\rm syn}$ to their optimal values at the
resonance ($D_3=4$, $\tau_{\rm syn}=35ms$) and compute the probability of observing a spike with rate $f_r$ for increasing $\Delta f$.  The results are plotted (in gray scale) in Fig. \ref{fig:frf1} as a function of
$f_1$, and show that the largest probability corresponds to spike rates following the prediction of relation
relation (\ref{eq:frf1}). Changing the noise intensity only obscures the observation of the spike density, but it
does not affect the agreement with the theoretical expression. In the bottom of Fig. \ref{fig:frf1} one can also see traces of less probable spikes, corresponding to a trivial subharmonic response of the system.

Figure \ref{fig:frf1} shows that the processing neuron emits spikes following Eq.~(\ref{eq:frf1}) for $k=2,3,4,5$. As
mentioned above, this relation is sustained by experimental data of pitch detection \cite{schouten}. Those experimental results indicate that equidistant tones in monoaural presentation do not produce constant pitch, similarly to what we observe in our binaural numerical experiments. We are not aware of binaural human experiments shifting the frequency components as in \cite{schouten}, which would be interesting to compare with our numerical predictions in Eq.~(\ref{eq:frf1}).

\section{Conclusions}

In this paper we demonstrate the phenomenon of ghost stochastic resonance in a neural circuit where two neurons
receive two components of a complex signal and their outputs drive a third neuron that processes the
information. The results show that the processing neuron responds preferentially at the ``missing fundamental''
frequency, and that this response is optimized by synaptic noise and by synaptic time constant. The processing
neuron is able to detect the coincident arrival of spikes from each of the input neurons, and this coincidence
detection is analogous to the linear interference of harmonic components responsible of the ghost response in
the single-neuron case \cite{dante1}. A brain structure candidate for this dynamics is the inferior colliculus,
which receives multiple inputs from a host of more peripheral auditory nuclei. Details of the physiology of this nucleus
are still uncertain, but enough evidence suggests that temporal and frequency representations of the inputs are
present in the spike timing of their neurons. Our results suggest that the neurons in this nucleus can exhibit the
dynamics described here, thus participating in the perception of binaural pitch. The main consequence of these
observations is that pitch information can be extracted mono or binaurally via the same basic principle, {\em i.e.}
ghost stochastic resonance, operating either at the periphery or at higher sensory levels.

\begin{acknowledgments}
We thank Dante R. Chialvo for guidance and useful comments on the manuscript.
We acknowledge financial support from MCyT-FEDER (Spain, projects BFM2002-04369 and BFM2003-07850), and by the
Generalitat de Catalunya. P.B. acknowledges financial support from the Fundaci\'on
Antorchas, Chile 300 (1098), Buenos Aires, Argentina.
\end{acknowledgments}

%\appendix

%\bibliography{ghostres}

\end{document}